\documentclass[twocolumn,aps,pre,showpacs,longbibliography]{revtex4-2}

\usepackage{tabularx}
\usepackage[latin1]{inputenc}
\usepackage{graphicx}
\usepackage{latexsym}
\usepackage{amssymb}
\usepackage{amsmath}
\usepackage{color}
\usepackage{tikz}
\usepackage{lipsum}
\usepackage{soul}
\usepackage{cancel}
\usepackage{enumitem}
\usepackage{hyperref}
\hypersetup{colorlinks=true,allcolors=blue}

\usepackage{amssymb}

\begin{document}
\def\d{{\rm d}}
\def\ex{{\rm e}}
\def\im{{\rm i}}
\def\J{{\bf J}}
\def\H{{\bf H}}
\def\E{{\bf E}}
\def\X{{\bf X}}
\def\M{{\bf M}}
\def\bfomega{{\boldsymbol{\omega}}}
\def\bfOmega{{\boldsymbol{\Omega}}}
\def\l{{\bf l}}
\def\k{{\bf k}}
\def\j{{\bf j}}
\def\e{{\bf e}}
\def\Emax{{E_c^{\operatorname{max}}}}
\def\iv{{\rm iv}}
\def\sign{{\rm sign}}
\def\Lcal{\mathcal{L}}
\def\bEcal{\boldsymbol{\mathcal{E}}}
\def\smalze{{\scriptscriptstyle{(0)}}}
\def\smalun{{\scriptscriptstyle{(1)}}}
\def\smaldu{{\scriptscriptstyle{(2)}}}
\def\smalm{{(m)}}
\def\smali{{\scriptscriptstyle{i}}}
\def\smalif{{\scriptscriptstyle{if}}}
\def\beq{\begin{eqnarray}}
\def\eeq{\end{eqnarray}}

\newcommand{\stc}[1]{\textcolor{magenta}{\st{#1}}}
\newcommand{\tc}[1]{\textcolor{magenta}{#1}}


\title{Ergodicity breaking and restoration in 
models of heat transport with microscopic reversibility}
\author{Piero Olla}
\thanks{Email address for correspondence: olla@dsf.unica.it}
\affiliation{ISAC-CNR and Istituto Nazionale di Fisica Nucleare, 
Section Cagliari, I---09042 Monserrato, Italy}

\begin{abstract}
The behavior of lattice models in which time reversibility is enforced at the level of
trajectories (microscopic reversibility) is studied analytically. Conditions for ergodicity 
breaking are explored, and a few examples of systems characterized by an additional conserved 
quantity besides energy are presented. All the systems are characterized by ergodicity
restoration when put in contact with a thermal bath, except for specific choices of the
interactions between the atoms in the system and the bath. The study shows that 
the additional conserved quantities return to play a role in non-equilibrium conditions. 
The similarities with the behavior of some mesoscale systems,
in which the transition rates satisfy detailed balance but not microscopic 
reversibility, are discussed.

\end{abstract}
\pacs{05.10.Gg,05.40.-a,05.10.Ln}
\maketitle

\section{Introduction}
\label{Introduction}
The derivation of equilibrium statistical mechanics starts in most textbooks from the 
assumption of microscopic Hamiltonian dynamics. The latter implies the conservation of phase space 
volume and that the equilibrium distribution of a macroscopic isolated 
system can only be a function of conserved quantities (Louville's theorem). Not all 
conserved quantities, however, fit the job. ``Good'' conserved quantities must have 
clear macroscopic content, which means that they are the result of a coarse-graining 
of microscopic degrees of freedom or that they fix the
value of macroscopic observables obtained by a coarse-graining procedure 
\cite{khinchin}.
Another condition is robustness, which means that if the system is in contact with a
thermal bath, the conserved quantity must continue to
play a role, say, by entering explicitly the Gibbs distribution. However, 
it is also possible that the additional conserved quantity reveals its presence only in 
out-of-equilibrium conditions, as it happens, e.g.,  with enstrophy---
the second moment of vorticity---in two-dimensional 
(2D) hydrodynamics \cite{kraichnan80,miller90}.

Understanding what makes a good conserved quantity requires diving into the guts of the 
coarse-graining procedure. The operation is hopeless for atoms and molecules obeying 
deterministic (or quantum) dynamics, except for specific interactions. For instance, 
including a collisional component in the dynamics, as in the ding-a-lin and ding-dong models 
\cite{casati84,prosen92}, has been shown to lead to break-up of KAM surfaces.
Similar results have been obtained by considering microscopic constituents
with an internal rotational dynamics \cite{wang04} or coupling the system to a fixed
substrate \cite{hu98}.

An alternative strategy is to replace atoms and molecules by microscopic agents obeying a
stochastic dynamics. It is the approach utilized in innumerable microscopic 
models, with applications ranging
from fluid dynamics 
\cite{succi01} to the study of transport processes 
\cite{kipnis82}, to the dynamics of systems
whose microscopic components are not molecules, such as granular media \cite{lipowski02}, 
crowds \cite{bruno11}, trees in a 
forest \cite{clar94}, or traffic on a highway \cite{oloan98}. 

Contrary to the aforementioned models, which base their
strength precisely on the insensitivity of the macroscopic processes
on the details of the microscopic interactions,
the focus of the stochastic models to be analyzed in the present paper 
is on the information destruction 
in the coarse-graining operation. The microscopic dynamics of the models is chosen accordingly.
In particular, full invariance under time reversal (microscopic reversibility) 
and energy conservation in the interactions are enforced at the microscale
to mimic the behavior of systems governed by a Hamiltonian dynamics. 

The proposed approach can be used to model mesoscopic fluctuation in heat-conducting media.
More interestingly, it provides a framework for including non-ergodic effects in 
generic stochastic systems.  The example of a medium in which local defects
constrain the structure of the heat currents and 
lead to the existence of an additional conserved 
quantity besides energy is discussed. Depending on the choice of parameters, an infinite hierarchy
of conservation laws is established, analogous to those for
vorticity in 2D ideal hydrodynamics. 

Putting the system in contact 
with a thermal bath typically restores ergodicity, and equilibrium is 
described by the standard Gibbs distribution for energy. The opposite
requires careful tuning of the interaction between the 
atoms in the bath and the system.
From the point of view of a description of the equilibrium state of the system, therefore, 
the new conserved quantity would not qualify as ``good''
The situation differs from that of systems such as, say, a two-dimensional Ising model below the
critical temperature, which remains non-ergodic when put in contact with a thermal bath, due
to the presence of free-energy barriers that become infinite in the thermodynamic limit
\cite{van_hemmen80,van_enter84}.

The new conserved quantity returns to play a role in non-equilibrium conditions.
When the system is put in contact with thermostats at different temperatures, it starts to absorb
part of the energy flowing from one thermostat to the other,
through a process similar to that occurring in 
mesoscopic ratchets \cite{sinitsyn07} and other mesoscopic systems 
\cite{filliger07,sokolov98,pietzonka16,dotsenko13}.

The paper is organized as follows. In Sec. II, the concept of microscopic reversibility is 
reviewed, and applied to the specific case of a  system with two sites.
The general formalism for the treatment of a system with an arbitrary number of sites and 
generic interactions is introduced in Sec. III. 
In Sec. IV, the simplest example of ergodicity breaking in a system with 
thee-atom interactions is presented. The interaction with a thermal bath is
discussed in Sec. V. Applications of the techniques to linear chains with two- and three-atom
interactions are illustrated in Secs. VI and VII. An extension of the analysis to non-equilibrium
conditions is presented in Sec. VIII.  Sec. IX is
devoted to the conclusions. Technical details are left in the appendices.

%

\section{Microscopic reversibility}
\label{Microscopic reversibility}
The present study focuses on a class of lattice models with a dynamics governed by random
energy exchange between the sites of the lattice without any role played by linear momentum
and other quantities describing the state of the system.  A Markovian dynamics is assumed.
Such models can then be seen as the overdamped limit of Hamiltonian
systems with a ``microcanonical'' stochastic component, of the kind described, e.g., in
\cite{kipnis82,basile06,lepri09}.

Microscopic reversibility can be expressed in terms of 
the transition PDF (probability density function) as
\beq
\rho(\E,t_f|\E',t_i)=\rho(\E',t_f|\E,t_i),
\label{microrev}
\eeq
where $t$ indicates time and $\E=(E_1,\ldots E_N)$ is the vector of the energies of the 
``stochastic atoms'' in the lattice.
The atoms are assumed to have a
positive energy spectrum to guarantee that the dynamics is bounded.

Note that the condition in Eq. (\ref{microrev}) is stronger than detailed balance 
(macroscopic reversibility)
\beq
\rho(\E,t_f|\E',t_i)\bar \rho(\E')=\rho(\E',t_f|\E,t_i)\bar \rho(\E),
\label{macrorev}
\eeq
where $\bar\rho$ is the equilibrium PDF. 
Equation (\ref{microrev}) implies that $\bar\rho$ can only be a function
of the total energy $E=\sum_kE_k$ and of the other conserved quantities
of the system (if present),
\beq
\bar\rho(\E)=g(E,\ldots).
\eeq
If $E$ is the only conserved quantity,
the microcanonical distribution is recovered.

\subsection{The two-atom system}
\label{The two-atom}
The microscopic reversibility condition Eq. (\ref{microrev}) is realized in the simplest
way in a two-atom system. A time-independent dynamics is assumed. After discretizing
time and energy at scales $\delta t$ and $\delta\hat E$,  
Eq. (\ref{microrev}) takes then the form, inspired by second quantization,
\beq
\rho(\E',t+\delta t|\E,t)
=\hat\Gamma f(\max(E_a,E_a'),\max(E_b,E'_b)),
\label{second quant}
\eeq
with the rule
\beq
E'_a=E_a\pm\delta\hat E,\qquad E'_b=E_b\mp\delta\hat E,
\label{rule}
\eeq 
to establish energy conservation, and the condition
\beq
f(E,0)=f(0,E)=0,\ \forall E,
\label{boundary 0}
\eeq
to guarantee $E_{a,b}>0$ at all times.
The parameter $\hat\Gamma$, which depends on the discretization, fixes the magnitude of the 
transition rate, and is introduced for book-keeping.

The similarity between Eq. (\ref{second quant}) and the transition rules in second quantization
becomes evident with the simple choice
\beq
f(\E)=E_aE_b.
\label{linear}
\eeq
Promoting $f$ to the role of second quantization operator would give, indeed,
$\hat f=\hat c_a^\dag\hat c_a\hat c_b^\dag\hat c_b$,
where $\hat c_{a,b}$ and $c^\dag_{a,b}$ are destruction and creation operators acting on states
$|N_a,N_b\rangle$, $N_{a,b}=E_{a,b}/\delta\hat E$. One finds immediately
$\langle N_a+1,N_b-1|\hat f|N_a,N_b\rangle=(N_a+1)N_b$
and $\langle N_a-1,N_b+1|\hat f|N_a,N_b\rangle=N_a(N_b+1)$, which reproduce Eqs.
(\ref{second quant}-\ref{boundary 0})

The dynamics of the system on the constant energy line $E_a+E_b=E$
is described naturally by the variable $E_{ab}^-=E_a-E_b$.
The moments of the increment
$\Delta E_a=-\Delta E_b=E_a(t+\Delta t)-E_a(t)$ read from
Eq. (\ref{second quant})
\beq
&&\langle\Delta E_a|\E\rangle=\frac{\Delta t\delta\hat E\hat\Gamma}{\delta t}
[f(E_a+\delta\hat E,E_b)
\nonumber
\\
&&\qquad-f(E_a,E_b+\delta\hat E)]
\simeq \frac{\Delta t\delta\hat E^2\hat\Gamma}{\delta t}\partial_{E^-_{ab}}f(\E),
\label{moment 1}
\\
&&\langle(\Delta E_a)^2|\E\rangle=\frac{\Delta t\delta\hat E^2\hat\Gamma}{\delta t}
[f(E_a+\delta\hat E,E_b)
\nonumber
\\
&&\qquad+f(E_a,E_b+\delta\hat E)]
\simeq \frac{2\Delta t\delta\hat E^2\hat\Gamma}{\delta t}f(\E).
\label{moment 2}
\eeq
From Eqs. (\ref{moment 1}) and (\ref{moment 2}) 
one obtains, by taking a continuum limit, the
stochastic differential equation (SDE) 
\beq
\dot E^-_{ab}=\Gamma\partial_{E_{ab}^-}f+\sqrt{2\Gamma f}\xi,
\label{SDE a due}
\eeq
where $\Gamma=2\hat\Gamma\delta\hat E^2/\delta t$,
$\langle\xi(t)\xi(0)\rangle=\delta(t)$, and the noise term is understood in the It\^o sense 
\cite{schuss}.
The associated Fokker-Planck equation reads
\beq
\dot\rho\equiv\partial_t\rho=\Gamma\partial_{E^-_{ab}}(f\partial_{E_{ab}^-}\rho),
\label{Fokker-Planck a due}
\eeq
which has a uniform distribution as a stationary solution.

\section{General formalism for the treatment of many-atom interactions}
In general, the dynamics of an $N$-atom system will be governed by interactions involving
an arbitrary number of atoms. Each interaction $l$ can be associated with a vector
$\J_l$ in the phase space of the system, such that 
the increment of $E^k$ produced by the interactions in the time interval $\delta t$ reads
\beq
\delta E^k= {J_l}^k\delta X^l, \qquad \delta X^l=0,\
\pm\delta\hat E 
\eeq
(the Einstein summation convention over repeated covariant and contravariant indices is
utilized throughout the section). 
For an $N$-atom system with $M$ conserved quantities
(counting the total energy), there will be at most $L=N-M$ independent interactions, which
will be here assumed to be also statistically independent,
$\rho(\E',t+\delta t|\E,t)=\prod_l\rho(\delta X^l|\E)$.

From the vectors $\J_l$, one can introduce curvilinear coordinates $X^k$ on the
hypersurface $\bfOmega$ of constant conserved quantities for the system
\footnote{In this framework, the energies $E^k$ 
are interpreted 
as cartesian coordinates, thus $E_k=E^k$ and
${J^k}_l=J^{kl}$.}. To obtain the new coordinates, one introduces dual vectors $\H_k$,
associating to vectors $\d\E$ components $\d X^l={H_k}^l\d E^k
=H^{kl}\d E_k$, such that
${J_l}^m\d X^l$ is the component $m$ of the projection of $\d\E$ on
$\bfOmega$. This is accomplished by imposing
${H_k}^l\d E^k=0$ for any vector $\d\E$ such that
${J_l}^k\d E_k=0$, and requiring
that $H$ and $J$ are one the inverse of the other on
$\bfOmega$, i.e.  ${J_l}^k{H_k}^m=\delta_l^m$.

Following a procedure analogous to that utilized for similar problems in classical mechanics
\cite{arnold}, it is possible to associate to interactions, generators
\beq
\hat\partial_{X^l}={J_l}^k\partial_{E^k},
\label{generator}
\eeq
such that $\delta f=\delta X^l\hat\partial_{X^l}f$ is the increment
of the function $f=f(\E)$ in the timestep $\delta t$. One readily verifies that
$\delta X^l\hat\partial_{X^l}f=\delta E^l\partial_{E^l}f$. Note that $\hat\partial_{X^l}
=\partial_{X^l}$ only if $L=N$.

The
conservation laws for the energy can be expressed in terms of generators as
$\hat\partial_{X^l}E=0$, $\forall l$, and identical relations will hold for the other
conserved quantities.
In the case of the two-atom system of Sec. \ref{The two-atom},
$\hat\partial_X=\partial_{E_{ab}^-}$ and $X=E_{ab}^-/2$.

The condition of microscopic reversibility is imposed as done in Eq.
(\ref{second quant}).
Define 
\footnote{
The relation ${E'}^k=E^k+\delta X^l{J_l}^k$ contains implicitly a maximum condition,
$\J_l=\J_l(\{\max(E_n,E_n')\})$. This is ultimately
responsible for the divergence term $(\partial_{E^k}{J_l}^k)$ in Eq. (\ref{moment 1 N}).
}
\beq
\rho(\delta X^m|\E)=\hat\Gamma f^\smalm(\{\max({E'}^k,E^k)\}),
\label{magilla}
\eeq
where ${E'}^k=E^k+\delta X^m{J_m}^k$ (no summation over $m$ here),
and impose 
\beq
E^l=0\Rightarrow f^\smalm(\E)=0,\ \forall l,m,
\label{boundary}
\eeq
which generalizes Eq. (\ref{boundary 0}) to the case of the generic $N$-atom system.
Indicate
\beq
f^{lm}=\delta^{lm}f^\smalm.
\eeq
Equation (\ref{moment 1}) becomes
\beq
\langle\delta X^m|\E\rangle&=&\hat\Gamma\sum_{\delta X^m}
\delta X^mf^\smalm(\{\max(E^k,{E'}^k)\})
\nonumber
\\
&\simeq&\hat\Gamma\sum_{\delta X^m} \delta X^m\delta X^l \partial_{E^k} ({J_l}^k f^{lm})
\nonumber
\\
&=&\hat\Gamma\delta E^2
\partial_{E^k} ({J_l}^k f^{lm}),
\nonumber
\\
&=&\Gamma[\hat\partial_{X^l}+(\partial_{E^k}{J_l}^k)] f^{lm} \delta t,
\label{moment 1 N}
\eeq
where use has been made of Eq. (\ref{generator}).
Proceeding in similar fashion for the increment correlation one gets
\beq
\langle\delta X^l\delta X^m|\E\rangle=2\Gamma f^{lm}\delta t.
\label{moment 2 N}
\eeq

In the following, only interaction rules that are independent of $\E$ will be considered,
which means that the vectors $\J_l$ and $\H_l$ are constant, and the $X^l$'s
are not curvilinear
coordinates (although, in general, they will remain not orthogonal,
$\d X^i={J^i}_k\d E^k\ne\d X_i={H^k}_i\d E_k$). Also,
the conserved quantities of the system  will be linear
combinations of the $E^k$'s, and the $\bfOmega$ hypersurfaces will be
hyperplanes.
The SDE for the system thus becomes
\beq
\dot X^l=\Gamma\hat\partial_{X^k}f^{lk}+\xi^l,
\quad\langle\xi^k(t)\xi^l(0)\rangle=2\Gamma f^{lk}\delta(t).
\label{SDE}
\eeq
and the associated Fokker-Planck equation will read
\beq
\dot\rho=\Gamma\hat\partial_{X^l}(f^{lk}\hat\partial_{X^k}\rho).
\label{Fokker-Planck}
\eeq
Extension of the formalism to the case of $\E$-dependent interaction rules
is provided for reference
in Appendix \ref{Treatment}.

\section{Ergodicity breaking in a three-atom system}
\label{Ergodicity breaking}
It is possible to show that in a system governed by binary interactions, energy is the only 
conserved quantity, provided any pair of atoms $l$ and $k$
can be connected by a path of binary interactions whose result is
a transformation $(E_l,E_k)\to(E_l\pm\delta\hat E,E_k\mp\delta\hat E)$.  Indeed, 
if the condition is satisfied,
the transformation will span at any given point $\E$
all the directions in the surface of constant $E$, and there will be
no regions in that surface not accessible from $\E$. In other words,
the system is ergodic.
\begin{figure}[h]
\begin{center}
\includegraphics[draft=false,width=3.2cm]{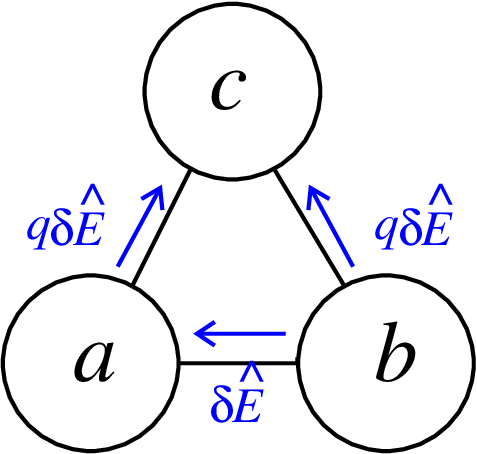}
\end{center}
\caption{Interaction rule for the three-atom system
}
\label{microfig1}
\end{figure}

Ergodicity breaking in a system that is not partitioned into isolated subsystems 
requires at least three-atom interactions. It is easy to verify
that the rule 
\beq
E_a&\to& E_a\pm (1-q)\delta\hat E,
\nonumber
\\
E_b&\to& E_b\mp(1+q)\delta\hat E,
\nonumber
\\
E_c &\to&
E_c\pm 2q\delta\hat E
\label{three-atom rule}
\eeq
(see Fig \ref{microfig1})
accomplishes the task.
A new conserved quantity is indeed present,
\beq
\omega=E_c+(E_b-E_a)q.
\label{omega}
\eeq
which
gives the degree of asymmetry of the energy distribution of the interacting atoms, and
will be here referred to, for lack of better names, as ``twist''.
The dynamics of the system is one-dimensional, and is described by the generator
\beq
\hat\partial_X
=2q\partial_{E_c}+(1-q)\partial_{E_a}-(1+q)\partial_{E_b},
\label{generator three-atom}
\eeq
corresponding to
\beq
X=\frac{2qE_c+(1-q)E_a-(1+q)E_b}{2(1+3q^2)}.
\label{correspondingly}
\eeq

It is possible to determine the range of variation of the different quantities describing
the system dynamics. One readily finds from the condition $E_l>0$ and Eq. (\ref{omega}),
\beq
-Eq<\omega<E,
\label{range q}
\eeq
and therefore also
\beq
\max(0,\omega-qE)<E_c<\omega+qE.
\label{range E_c}
\eeq

Of special interest is the range $q\ll 1$, in which the three-atom dynamics approximates
that of the two-atom system.
For $\omega/E=O(q)$, Eq. (\ref{range E_c}) implies $E_c/E=O(q)$, and therefore $E\simeq E_a+E_b$, 
$\omega\simeq (E-2E_a)q+E_c$. For $\omega/E=O(q)$ with $\omega>qE$, the range of variation of the vector 
$(E_a,E_b)$
is along the segment in the $(E_a,E_b)$ plane with endpoints $(0,E)$ and $(E,0)$, which
reproduces the dynamics
of the two-atom system. If instead $-qE<\omega<qE$, the vector
$(E_a,E_b)$ will vary along the segment with endpoints $((E-\omega/q)/2,E)$ and $((E+\omega/q)/2,0)$. 
Finally, 
if $\omega=\chi E$ with $q\ll\chi<1$, $\chi-(1-\chi)q<E_c/E<\chi+(1+\chi)q$, and
the vector $(E_a,E_b)$ varies along the segment with endpoints $(0,(1-\chi)E)$ and
$((1-\chi)E,0)$; the behavior of the two-atom system is retrieved again.

\section{Interaction with a thermal bath}
\label{Interaction with a thermal bath}
Following standard practice, interaction with a thermal bath is modeled by imagining 
the system $(A)$ as a part of a much larger isolated system, in which the state of
the remnant (the bath, $B$) is known only in an average sense.
In principle, system $AB$ could  be a homogeneous unit, possibly with 
multi-atom and long-range interactions. More typically, $A$ and $B$ are distinct units interacting
by short-range forces. Assuming binary interactions 
with strength determined by Eq. (\ref{linear}), leads to a Fokker-Planck equation 
for $\rho(\E,t)\equiv\rho(\E^A,\E^B,t)$ in the form 
\beq
\dot\rho&=&\Gamma_A\sum_k\hat\partial_{X_k}(f_k\hat\partial_{X_k}\rho)
\nonumber
\\
&+&\Gamma_{AB}\sum_{kl}
\partial_{E_{kl}^-}(E^A_kE^B_l\partial_{E_{kl}^-}\rho)+\Lcal^\dag_B\rho,
\label{Fokker-Planck AB}
\eeq
where $kl$ labels the interacting pairs in $A$ and $B$ and $\Lcal^\dag_B\rho$ accounts
for the internal dynamics of $B$. 

Indicate with $T$ the bath temperature and set the Boltzmann constant $=1$, in
such a way that $\langle E_l^B\rangle=T$.
Integrating Eq. (\ref{Fokker-Planck AB}) over $[\d E^B]=\prod_l\d E_l^B$, one obtains
after a little algebra the Fokker-Planck equation for the marginal $\rho_A$
\beq
\dot\rho_A&=&\Gamma_A\sum_\alpha\hat\partial_{X_\alpha}(f_\alpha\hat\partial_{X_\alpha}\rho_A)
\nonumber
\\
&+&\Gamma_{bath}\sum_k
\partial_{E_k}[E_k(T\partial_{E_k}+1)\rho_A],
\label{Fokker-Planck bath}
\eeq
where $\Gamma_{bath}=N_{ab}\Gamma_{ab}$ and $N_{ab}$ is the number of interacting atom 
pairs in $A$ and $B$.
The atoms in $A$ interacting with the bath will then obey the SDE
\beq
\dot E_k+\Gamma_{bath}(E_k-T)=\xi_k+\dot E_k^A,
\label{SDE bath}
\eeq
where
\beq
\langle\xi_k(t)\xi_j(0)\rangle=2\Gamma_{bath}TE_k\delta_{jk}\delta(t),
\eeq
and $\dot E_k^A$ accounts for the interactions of atom $k$ with the other atoms in $A$.
Averaging over the degrees of freedom in $B$ replaces the interaction between atoms
in $A$ and $B$, originally in the symmetric form $\Gamma_{AB}(E_k-E_l)$, with 
the relaxation term $\Gamma_{bath}(E_k-T)$, which 
breaks microscopic reversibility, and establishes an arrow of time at the level
of trajectories. Note that
Eq. (\ref{SDE bath}) is the same kind of energy SDE that would be produced in systems whose 
interaction with the heat bath is described by a linear Langevin equation in configuration 
space (in phase space if the system is underdamped).

If $E_A$ is the only conserved quantity for system $A$, the stationary solution of 
Eq. (\ref{Fokker-Planck bath}) will be constant on the surfaces at fixed $E^A$. The
condition
$\rho(E_k)=T^{-1}\exp(-E_k/T)$ for the atoms in $A$ which interact with $B$ then implies that
the stationary distribution has the Gibbs form
\beq
\rho_A=Z_A^{-1}\exp(-E^A/T).
\label{Gibbs}
\eeq
The situation with more than one conservation law is less clear
(a trivial 
example is that of $B$ interacting with a portion of $A$ isolated from the
rest of the system), and one must analyze the situation on a case-by-case basis.

\section{Simple atom chain}
\label{The atom chain}
\begin{figure}[h]
\begin{center}
\includegraphics[draft=false,width=6cm]{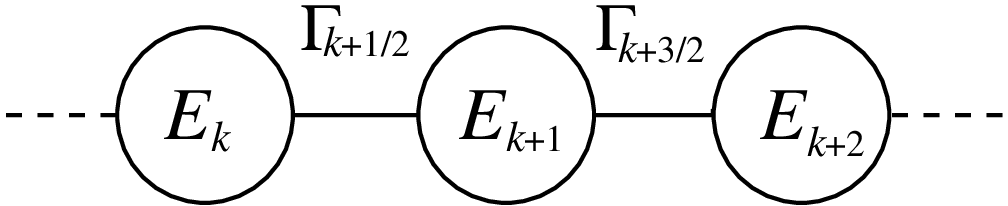}
\end{center}
\caption{Sketch of the simple atom chain.
}
\label{microfig2}
\end{figure}
The two-atom system in Sec. \ref{The two-atom} can be used as a building block 
for cubic lattices in an arbitrary number of dimensions. 
The simplest realization is that of a linear chain.
The absence of a locally conserved linear momentum makes one expect that anomalous behaviors
observed in low dimensional systems \cite{narayan02,lepri09} are not an issue here
(see \cite{basile06} for another example of
microscopically reversible stochastic model which includes local momentum conservation).
The dynamics of the linear chain in Fig. \ref{microfig2} could then be expected to be
representative of that of higher dimensional lattice models \cite{dhar08}. 

Assume for definiteness a linear relaxation dynamics, as 
described by Eq. (\ref{linear}). 
The Fokker-Planck equation for the isolated atom chain will read
\beq
\dot\rho=\sum_{k=0}^{N-1}\Gamma_{k+1/2}
\partial_{E^-_{k,k+1}}(E_kE_{k+1}\partial_{E^-_{k,k+1}}\rho),
\label{Fokker-Planck chain}
\eeq
corresponding to the SDE
\beq
\dot E_k&=&
\Gamma_{k-1/2}(E_{k-1}-E_k)
\nonumber
\\
&+&\Gamma_{k+1/2}(E_{k+1}-E_k)+\xi_k,
\label{SDE chain}
\eeq
where
\beq
\langle\xi_k(t)\xi_k(0)\rangle&=&
2E_k(\Gamma_{k+1/2}E_{k+1}+\Gamma_{k-1/2}E_{k-1})\delta(t),
\nonumber
\\
\langle\xi_k(t)\xi_{k+1}(0)\rangle&=&-2\Gamma_{k+1/2}E_kE_{k+1}\delta(t),
\label{noise chain}
\eeq
and $\langle\xi_k\xi_j\rangle=0$ for $|k-j|>1$. To extend the expressions to the case of dimension
$D$ generic, it is sufficient to
make in Eqs. (\ref{Fokker-Planck chain}-\ref{noise chain}) the substitutions
$k\to\k=(k_1,\ldots k_D)$, $k+1\to \k+\e_\alpha$, $e_{\alpha\beta}=\delta_{\alpha\beta}$,
$\Gamma_{k+1/2}\to \Gamma_{\k+\e_\alpha/2}$, 
and sum over $\alpha$ in the right hand side of Eqs. (\ref{Fokker-Planck chain})
and (\ref{SDE chain}).

The deterministic terms in Eq. (\ref{SDE chain}) 
have the same form as in any models of heat
transport based on random energy exchange among sites in the lattice.
However, some differences between  models
exist at the level of interpretations, especially if the chain is spatially inhomogeneous.

Consider the case of the zero-range model \cite{evans05}. 
By construction, the transition rate $\Gamma$
in that model is associated with the sites rather than the links, which leads to spurious energy
currents from sites 
with larger $\Gamma$ to sites with smaller $\Gamma$.  
Different transition rates
must be imposed to the left and the right of each site to cancel the currents, 
which makes the dynamics
of the individual atoms dependent on the geometry of the chain. This does not occur
in the present model, in which the transitions take place on the links rather than at the sites
of the lattice.

A more substantial difference concerns the noise. Consider again the zero-range model.
The transition rate and, hence, also the noise amplitude for the zero-range model 
are linear in $E_k$, while that of
the present model is quadratic in $E_k$ 
[see Eq. (\ref{noise chain})]. By dimensional consistency, 
the noise amplitude in the zero-range model satisfies
\beq
\langle\xi\xi\rangle\sim\Gamma E\delta\hat E,\quad\operatorname{zero-range},
\eeq
and vanishes for $\delta\hat E\to 0$, while that of the present model remains finite in the limit.
Thus, a zero-range model could not be utilized to describe
mesoscale fluctuations in heat transport, while Eqs. 
(\ref{Fokker-Planck chain}-\ref{noise chain}) in principle can.

\subsection{Non-equilibrium steady states}
Suppose the chain is in contact at its endpoints with thermal baths at temperatures $T_{0,N}$.
Consider first the case of a system at equilibrium with $T_0=T_N=T$.
The conditions for Gibbs statistics  discussed in 
Sec. \ref{Interaction with a thermal bath} are satisfied, hence
\beq
\bar E_l=T,\qquad
\langle\tilde E_l\tilde E_k\rangle=T^2\delta_{lk},
\eeq
where $\bar E_l\equiv\langle E_l\rangle$ and $\tilde E_l=E_l-\bar E_l$.

For $T_N\ne T_0$ a non-equilibrium stationary state (NESS) is established, with a linear
profile for the mean energy 
\beq
\bar E_k=\bar E_0+\bar E'k,
\label{E'}
\eeq
where $\bar E_0=T_0$ and $\bar E'=(T_N-T_0)/N$.

Non-equilibrium fluctuations are better analyzed by decomposing
\beq
\tilde E_k=\tilde E_k^{le}+\tilde E_k^r,
\label{decomp}
\eeq
where 
$\tilde E^{le}_k$ is part of the fluctuation that describes local thermal equilibrium,
\beq
\langle\tilde E^{le}_l\tilde E^{le}_k\rangle=\bar E_l\delta_{lk}.
\label{condit1}
\eeq
The remainder $\tilde E^r$ is assumed uncorrelated with $\tilde E^{le}$,
in such a way that
\beq
\langle\tilde E_l\tilde E_k\rangle=
\langle\tilde E^r_l\tilde E^r_k\rangle:=C_{lk},\quad l\ne k.
\eeq

\begin{figure}[h]
\begin{center}
\includegraphics[draft=false,width=7cm]{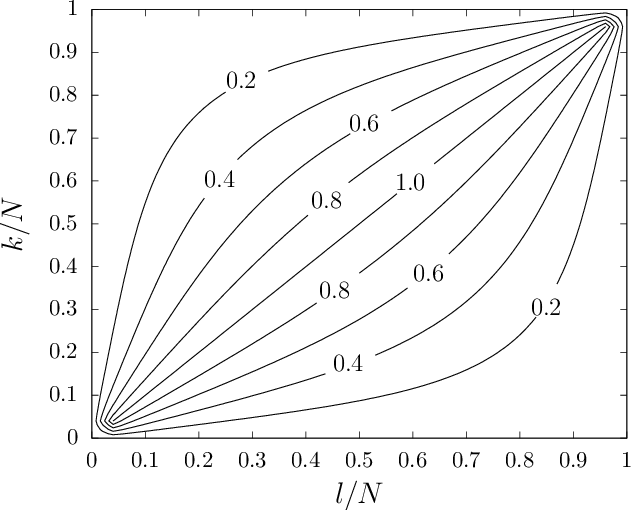}
\end{center}
\caption{Contour plot of the normalized non-equilibrium correlation component 
$\hat C_{kl}=3C_{kl}/(2{E'}^2)$.
}
\label{microfig3}
\end{figure}

Consider the case of a spatially homogeneous chain for simplicity.
A major question is whether the fluctuations have a long correlation component. A preliminary
answer can be obtained by looking at the decay of $C_{lk}$ for $0\ll l\ll N$, similarly for
$k$, with
$1\ll |l-k|\ll N$. In this range, $C_{lk}\simeq C(l-k)$ and the result extends to higher dimensions,
$C_{\k\l}\simeq C(\k-\l)$.
One can verify from Eqs. 
(\ref{Fokker-Planck chain}-\ref{noise chain}) that for $1\ll |\l-\k|$ the fluctuation correlation
obeys a discrete Laplace equation
\beq
\nabla^2_{\l\k}C_{\l\k}=0,
\label{Laplacian 2D}
\eeq
where $\nabla^2_{\l\k}=\sum_{\j=\k,\l}\sum_{\alpha=1}^D\partial_{j_\alpha}^2$,
with $\partial_{j_\alpha}f_\j=f_{\j+\e_\alpha}-f_\j$
(see Appendix \ref{Fluctuation spectrum}). For $|\l-\k|\ll N$, the derivatives along 
$\l+\k$ do not contribute to $\nabla^2_{\l\k}$, which, from $2D$-dimensional, 
becomes a $D$-dimensional Laplacian, $\nabla^2_{\l\k}\to \nabla^2_{\l-\k}$. Equation
(\ref{Laplacian 2D}) thus becomes $\nabla^2_{\l-\k} C(\l-\k)=0$, leading to the behaviors
\beq
C(\l-\k)\sim \begin{cases}
|\l-\k|,\qquad &D=1,
\\
\ln|\l-\k|,&D=2,
\\
|\l-\k|^{-1},&D=3.
\end{cases}
\eeq
An explicit calculation carried out in Appendix \ref{Fluctuation spectrum} for $D=1$ 
confirms the result and gives the correlation fluctuation correction on the diagonal
\beq
C_{ll}=\frac{2}{3}{E'}^2.
\label{BCdiag}
\eeq
The profile of the function $C_{kl}$ in Fig. \ref{microfig3}
shows  that for $D=1$ the energy fluctuations, indeed, have a 
long-correlation component analogous to
those for momentum and displacement in low-dimensional systems with linear momentum conservation.
It is interesting to note that momentum and displacement fluctuation have zero correlation
in the case of the purely viscous chain \cite{olla23}, which shares with the present system
a linear profile for the mean energy.
Contrary to the case of a $D=1$ system
with linear momentum conservation, however, 
the amplitude of the long-range correlations vanishes like $N^{-2}$ for $N$ large and 
fixed value of the temperature gap $T_N-T_0$.

\section{The double chain}
\label{The double chain}
\begin{figure}[h]
\begin{center}
\includegraphics[draft=false,width=8cm]{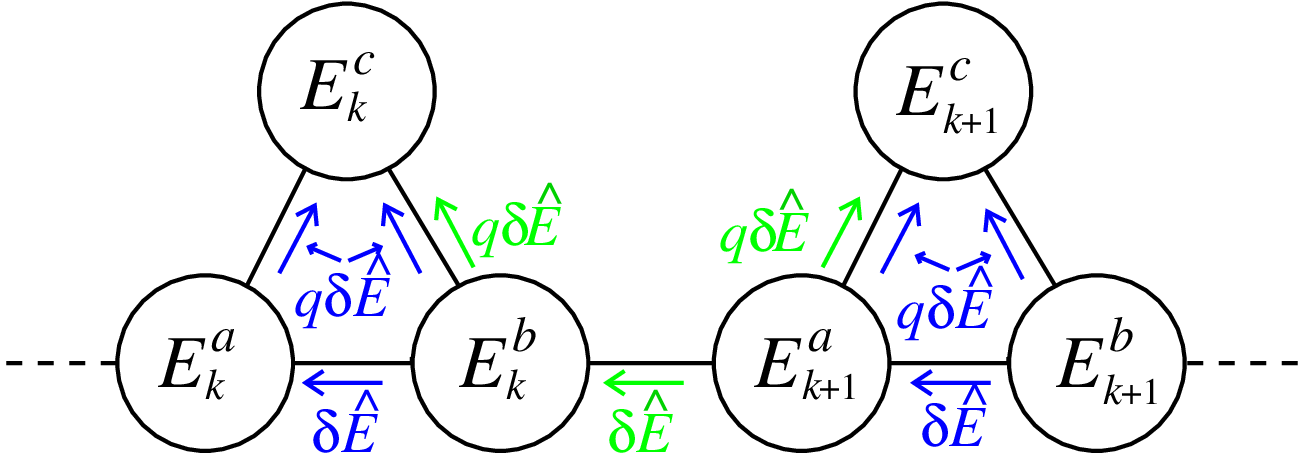}
\end{center}
\caption{The double chain geometry. Blue for interaction $X_k$ 
[Eqs. (\ref{three-atom rule}) and (\ref{generator three-atom})];
green for $X_{k,k+1}$ [Eqs. (\ref{rule 2chain}) and (\ref{generator 2chain})].
}
\label{microfig4}
\end{figure}
By joining several three-atom units of the kind discussed 
in Sec. \ref{Ergodicity breaking} together,
one can build a linear chain endowed with multiple conservation laws. 
A possible geometry is that of Fig. \ref{microfig4}, in which the three-atom units
form ``stochastic molecules'', whose internal dynamics is described by
Eqs. (\ref{three-atom rule}) and (\ref{generator three-atom}).  The interactions between
the molecules, instead, are taken to obey the rule
\beq
E^b_k&\to&E^b_k\pm(1-q)\delta\hat E,
\nonumber
\\
E^a_{k+1}&\to&E^a_{k+1}\mp(1+q)\delta\hat E,
\nonumber
\\
E^c_k&\to&E^c_k+q\delta\hat E,
\quad
E^c_{k+1}\to E^c_{k+1}+q\delta\hat E,
\label{rule 2chain}
\eeq
corresponding to the generator
\beq
\hat\partial_{X_{k,k+1}}&=&q(\partial_{E_k^c}+\partial_{E_{k+1}^c})
\nonumber
\\
&+&(1-q)\partial_{E_k^b}
-(1+q)\partial_{E_{k+1}^a}.
\label{generator 2chain}
\eeq
The Fokker-Planck equation for the isolated chain will read (spatial homogeneity 
assumed)
\beq
\dot\rho=
\Gamma\sum_k[\hat\partial_{X_k}(f_k\hat\partial_{X_k}\rho)
+\hat\partial_{X_{k,k+1}}
(f_{k,k+1}\hat\partial_{X_{k,k+1}}\rho)],
\label{Fokker-Planck 3-chain}
\eeq
where the $X_k$'s account for the interactions within the stochastic molecules.

The system conserves both total energy $E=\sum_k E_k=\sum_{k,\alpha}E_k^\alpha$
and total twist $\omega=\sum \omega_k$. The latter conservation law 
could be extended to
any function $H(\bfomega)=\sum_kh(\omega_k)$ by replacing the continuous
dynamics in Eqs. (\ref{rule 2chain}) by jumps
\beq
(\omega_k,\omega_{k+1})\to (\omega_k',\omega'_{k+1})=(\omega_{k+1},\omega_k),
\label{jump}
\eeq
a behavior akin to that of vorticity in 
2D flows \cite{salmon88}. Such a dynamics could
be realized trivially by the exchange interaction 
$(\E_k,\E_{k+1})\to (\E'_k,\E'_{k+1})=(\E_{k+1},\E_k)$,
$\E_k\equiv (E_k^a,E_k^b,E_k^c)$.
Energy exchange between molecules could still be incorporated into the dynamics without
affecting their twist, e.g., by the rule
\beq
E^a_k,E^b_k&\to&E^a_k\pm\delta\hat E,E^b_k\pm\delta\hat E,
\nonumber
\\
E^a_{k+1},E^b_{k+1}&\to&E^a_{k+1}\mp\delta\hat E,E^b_{k+1}\mp\delta\hat E,
\label{altrule}
\eeq
and the extended set of conservation laws for  $\bfomega$ would be preserved.

\subsection{Stationary solution}
\label{Stationary solution}
Back to the continuous dynamics of Eqs.
(\ref{three-atom rule},\ref{generator three-atom},\ref{rule 2chain}-\ref{Fokker-Planck 3-chain}), 
progress in the analysis is possible for small $q$ by adopting a
multiscale approach \cite{bender}.
Inspection of Eqs. (\ref{generator three-atom}) and
(\ref{generator 2chain}-\ref{Fokker-Planck 3-chain})
suggests the ansatz
\beq
\rho(\E,t)=g(\E,q\E^a,q\E^b;t),
\eeq
where $\E^\alpha=(E^\alpha_1,\ldots E^\alpha_N)$, and one assumes that both $f_k$ and $f_{k,k+1}$
are $O(q^0)$.
To lowest order in $q$, Eq. 
(\ref{Fokker-Planck 3-chain}) yields, at stationarity,
\beq
f_k\partial_{E^{ab-}_k}\bar\rho=0,\qquad
f_{k,k+1}\partial_{E^{ba-}_{k,k+1}}\bar\rho=0,
\label{O(q^0)}
\eeq
where $E^{ba-}_{k,k+1}=E^b_k-E^a_{k+1}$, and, as before, $E^{ab-}_k=E^a_k-E^b_k$.
Equation (\ref{O(q^0)}) implies
\beq
\bar\rho(\E)=\exp[-(E_a+E_b)/T_{ab}]\hat g(\E^c,q\E^a,q\E^b),
\label{rho O(q^0)}
\eeq
pointing to the fact that to
lowest order in $q$, the lower and upper portion of the chain decouple, with the lower
portion behaving as a simple chain governed by binary interactions.

Going to the next order, one finds, again at stationarity,
\beq
f_k\Big[\partial_{E^c_k} +\frac{1}{2q}\partial_{E_k^{ab-}} +\frac{1}{T_{ab}} \Big]\hat g=0,
\nonumber
\\
f_{k,k+1}\Big[\partial_{E^c_k}+\partial_{E^c_{k+1}} +\frac{1}{q}\partial_{E_{k,k+1}^{ba-}}
 +\frac{2}{T_{ab}} \Big]\hat g=0,
\label{O(q)}
\eeq
which imply $\partial_{E_k^{ab-}}\hat g=\partial_{E_{k,k+1}^{ba-}}\hat g$, and therefore also
\beq
\hat g=
\exp\Big\{-\sum_k\frac{(1+\Phi)E^c_k-q\Phi E^{ab-}_k}
{T_{ab}}\Big\},
\eeq
where $\Phi$ is an arbitrary constant. The equilibrium statistics of a 
portion $A$ of an isolated chain 
with fixed values of $E$ and $\omega$ will thus be described by
a dual-temperature Gibbs distribution
\beq
\bar\rho(\E^A)=Z_A^{-1}\exp\Big[-\frac{E^A}{T}-\frac{\omega^A}{T_\omega}\Big],
\label{Gibbs 2chain}
\eeq
where $T=T_{ab}$ and $T_\omega=T_{ab}/\Phi$.
Dual-temperature systems have been studied indeed in a non-equilibrium context 
(see e.g. \cite{dotsenko13}). Equation (\ref{Gibbs 2chain}) provides an example
of the same phenomenon for systems at equilibrium (see \cite{miller90} for
a fluid mechanics example).
The canonical
and microcanonical distributions description are equivalent for the system under consideration.
It is possible to 
verify that including in the dynamics
a twist-preserving interaction like the one in Eq. (\ref{altrule}),
does not modify the result.  The standard form of the Gibbs distribution is 
recovered for $\Phi=0$, in which case $\langle \omega_k\rangle=T$.

\subsection{Interaction with a thermal bath and ergodicity restoration (not always)}
One may wonder whether
the dual-temperature form of the Gibbs distribution [ Eq. (\ref{Gibbs 2chain})]
is maintained when the system is in contact with a thermal bath. The answer  
turns out to be that it depends. 
Consider a situation in which only the endpoints of the 
chain are thermostatted,
as described by the Fokker-Planck equation
\beq
\dot\rho=(\Lcal^\dag_X+\Lcal_{bath}^{ab\dag}+\Lcal_{bath}^{c\dag})\rho,
\label{Fokker-Planck 2bath}
\eeq
where
\beq
\Lcal_{bath}^{ab\dag}\rho=\Gamma_{bath}^{ab}\Big\{
\partial_{E_0^a}[E^a_0(T^a_0\partial_{E_0^a}+1)\rho]
\nonumber
\\
+ \partial_{E_N^b}[E^b_N(T^b_N\partial_{E_N^b}+1)\rho]
\Big\},
\label{Fokker-Planck ab}
\\
\Lcal_{bath}^{c\dag}\rho=\Gamma_{bath}^c\Big\{
\partial_{E_0^a}[E^c_0(T_0^c\partial_{E_0^c}+1)\rho]
\nonumber
\\
+ \partial_{E_N^b}[E^c_N(T_N^c\partial_{E_N^c}+1)\rho]
\Big\},
\label{Fokker-Planck c}
\eeq
and $\Lcal^\dag_X\rho$ accounts for the internal dynamics of the chain, 
[see Eq. (\ref{Fokker-Planck 3-chain})]. Take for simplicity 
$\Gamma\gg\Gamma_{bath}^{ab},\Gamma_{bath}^c$, in such a way that the equilibration with the
thermal baths is slow and the chain can be assumed at all times 
in equilibrium with itself, as 
described by Eq. (\ref{Gibbs 2chain}).

Consider first the case in which only the endpoints of the lower chain are in contact with the
thermal baths, $\Gamma_{bath}^c=0$, and set $T_0^a=T_N^b=T$.
Substituting Eq. (\ref{Gibbs 2chain}) into Eqs.
(\ref{Fokker-Planck 2bath}) and (\ref{Fokker-Planck ab}), and exploiting Eq. (\ref{omega}), 
yields
\beq
\langle\dot E\rangle=O(\Gamma_{bath}^{ab}Tq^3),
\quad
\langle\dot\omega\rangle\simeq 2q^2\Phi\Gamma_{bath}^{ab}T.
\label{dot omega}
\eeq
Equation (\ref{Gibbs 2chain}) implies
$\langle \omega\rangle=N\langle \omega_k\rangle= NT/(1+\Phi)+O(NTq^2)$, 
in which case the second of Eq. (\ref{dot omega})
becomes
\beq
\langle\dot \omega\rangle=2q^2T\Gamma_{bath}\Big(\frac{NT}{\langle \omega\rangle}-1\Big),
\eeq
which describes the relaxation of the system statistics towards its standard 
one-temperature form, in which $\langle \omega\rangle=NT$.
Putting the endpoints of the lower portion of the double-chain in contact with 
a thermal bath is thus sufficient to restore ergodicity in the chain.

What if the previous operation is carried out with the  upper portion of the chain?
Repeating the same operations leading to Eq. (\ref{dot omega}) starting this time 
for $\Gamma_{bath}^{ab}=0$
from Eqs. (\ref{Fokker-Planck 2bath}) and (\ref{Fokker-Planck c}) and setting $T_0^c=T_N^c=T^c$,
yields
\beq
\langle\dot E\rangle=\langle\dot\omega\rangle=2\Gamma_{bath}^c\Big[T^c-\frac{T}{1+\Phi}\Big].
\eeq
Withing this arrangement, ergodicity is in general not restored.

The results are easy to understand in the case of a single three-atom unit. 
The relaxation of the system toward a zero twist regime is associated with a 
current from $a$ to $b$. This current is established naturally
when atoms $a$ and $b$ are in contact with thermal baths at temperarature $T$, and only ceases
when $\Phi=0$ [from 
(\ref{Gibbs 2chain}), $\langle E_{a,b}\rangle\simeq (1\pm q\Phi)T\ne T$]. On the
contrary, the only current that exists
when just atom $c$ is thermostatted originates from the internal equilibration of
the chain; it stops when $T=T^c(1+\Phi)$, which is not in general a $\langle\omega\rangle=NT$ 
single-temperature state.

\section{Non-equilibrium steady state for the three-atom system}
The most conspicuous characteristic of the three-atom system is its lack of reflection invariance.
The analysis in Sec. \ref{The double chain} shows that 
lack of reflection invariance in a double chain is a microscopic property without a
macroscopic counterpart at equilibrium unless the system is isolated or 
specific interactions are assumed with the surroundings. In other words,
twist is 
a ``bad'' conserved quantity, not contributing information on the macroscopic
state of the system at equilibrium.

The situation changes dramatically in non-equilibrium conditions. 
Consider a single three-atom unit,
with atoms $a$ and $b$ in contact with thermostats at 
temperatures $T_a\ne T_b$. The interaction rule in Eq. (\ref{three-atom rule}) dictates that
if $T_a>T_b$, energy will flow out of atom $c$ until $E_c\to 0$.
At this point, the (mean) heat flow from $a$ to $b$ 
vanishes, and the system will act as a thermal insulator. 
Conversely, if $T_a<T_b$, a heat flow from $b$ to $a$ is established, associated with 
a possibly unbounded growth of $E_c$.

\subsection{Stationary state}
\label{Vanishing}
Consider again a small $q$ regime, and assume $\Gamma_{bath}\gg\Gamma$, in such a way that
\beq
\rho(E_a,E_b|E_c)\simeq\frac{1}{T_aT_b}\exp\Big(-\frac{E_a}{T_a}-\frac{E_b}{T_b}\Big).
\label{NESS0}
\eeq
The Fokker-Planck equation for the system is in the form 
[compare with Eq. (\ref{Fokker-Planck bath})]
\beq
\dot\rho&=&\Gamma\hat\partial_X[f_X\hat\partial_X\rho]
\nonumber
\\
&+&\Gamma_{bath}\sum_{\alpha=a,b}
\partial_{E_\alpha}[E_\alpha(T_\alpha\partial_{E_\alpha}+1)\rho].
\label{Fokker-Planck NESS}
\eeq
Consider first the case in which the system behaves as a thermal insulator
One expects that a stationary state with vanishing heat flow 
will satisfy detailed balance \cite{ohga23}, even though it is not, stricly speaking, a
thermal equilibrium state.
Substituting Eq. (\ref{NESS0}) into Eq. (\ref{Fokker-Planck NESS}) and imposing 
zero current yields
\beq
\hat\partial_X\bar\rho=2q
(\partial_{E_c}+1/\bar E_c^s)\bar\rho=0,
\label{reduced}
\eeq
where
\beq
\bar E^s_c\simeq T\Big(1-\frac{\Delta T}{2qT}\Big)^{-1}
\label{E^s}
\eeq
and 
\beq
T_a=T-\Delta T,\qquad T_b=T.
\eeq
A stationary solution only exists for $\Delta T<2qT$, 
\beq
\bar\rho(E_c)=(1/\bar E_c^s)\exp(-E_c/\bar E_c^s).
\label{rho(E_c)}
\eeq
The system behaves in this case 
like a heat battery, in which $\bar E_c^s$ plays the role of maximum
charge. The process of ``battery charging'' is accompanied by the system
mean twist deviating  from its equilibrium value $\bar\omega=T$ and taking the role
of a sort of effective temperature for the non-equilibrium system \cite{dotsenko13,cerasoli18}.

One would expect that only non-stationary solutions exist
for $\Delta T\ge 2qT$, corresponding to growth without bound of $\bar E_c$.
The situation, however, is more complex. 
\begin{enumerate}[label=\roman*.]
\item
The system will behave as a heat conductor if, for large $E_c$,
$T^{-2}f_X$ tends to a constant $O(1)$ limit.
One would have in this case an example of a microscopic heat diode, which conducts
heat for $\Delta T>2qT$ and behaves as an insulator otherwise \cite{li04}. If $q$ vanishes at large $E_c$, 
the growth of $\bar E_c$ will stop; otherwise, it will continue without a bound.
\item
If $T^{-2}f_X$ vanishes at some $E_c$, the system will behave as an insulator,
irrespective of the sign of $\Delta T-2qT$.
\item
If $T^{-2}f_X$ increases monotonically with $E_c$, the condition $\Gamma\ll\Gamma_{bath}$ 
leading to Eq. (\ref{NESS0}) must be replaced by $\Gamma T^{-2}f_X\ll\Gamma_{bath}$, which
is going to be violated at sufficiently large $E_c$. When this happens, one must make
in Eq. (\ref{E^s}) the replacement
$\Delta T\to \Delta T_{eff}=\bar E_b-\bar E_a\sim \Gamma f_X\Delta T/(T^2\Gamma_{bath})$.
The growth of $\bar E_c$ will cease when $\Delta T_{eff}=2qT$,
at which point the heat flow vanishes (the same will
occur for $\Delta T<2qT$, when the value of $E_c^s$ predicted by Eq. (\ref{E^s}) is too large).
\end{enumerate}
The mechanism responsible for the diode effect described above is the nonlinearity of 
the dynamics required for the saturation of the normalized transition rate $T^{-2}f_X$,
coupled with the non-equilibrium nature of the process. 
The last condition is required to break the time reflection symmetry of the dynamics \cite{li21}.
Note that the first condition is violated in the initial phase of
the process, when $E_c$ is small, in which case it is possible to show that the 
heat flow is independent of the sign of $T_a-T_b$. 

The same diode effect could be realized in the case of a two-atom system by assuming an 
asymmetric form of the transition rate, say, $f(E_a=E',E_b=E-E')>f(E-E',E')$ for $E'>E/2$.
It is possible to see, indeed, that the heat flow from $a$ to $b$ in an asymmetric two-atom
system is larger in magnitude for $T_a>T_b$ than for $T_a<T_b$ (compare with 
the result in \cite{li04}). The only condition that must be satisfied is that
the temperature gap $T_a-T_b$ is non-negligible.

\subsection{Example of time-dependent solution}
More precise information on the non-equilibrium behavior of the three-atom system
can be obtained if 
a specific form for the interaction strengths is selected, e.g.
\beq
f_X=E_aE_bE_c/E_r,
\label{f_X}
\eeq
where $E_r$ is a fixed energy scale. Breakup of the approximation 
$\Gamma_{bath}\gg\Gamma f_X/T^2$ will take place for $E_c\sim E_r\Gamma_{bath}/\Gamma$.
However, 
if $E_r\gtrsim T$ and the system starts in a condition in which $E_c\sim T$, the
condition $\Gamma_{bath}\gg\Gamma f_X/T^2$ will be satisfied in the initial transient.
In this case, one can
substitute Eq. (\ref{NESS0}) and (\ref{f_X}) into Eq. (\ref{Fokker-Planck NESS}) 
and the result is
\beq
\dot\rho=
\frac{4q^2\Gamma T^2}{E_r}\partial_{E_c}[E_c(\partial_{E_c}+1/\bar E^s_c)\rho].
\label{FPNESS}
\eeq
From here, one gets evolution equations for $\bar E_c$
\beq
\dot{\bar E}_c=\frac{\bar E_c^s-\bar E_c}{\tau},\quad
\tau=\frac{E_r\bar E_c^s}{4q^2\Gamma T^2}
\label{tau}
\eeq
and for the fluctuations
\beq
\frac{\d}{\d t}\langle\tilde E_c^2\rangle
=\frac{1}{\tau}(\bar E_c^s\bar E_c-\langle\tilde E_c^2\rangle),
\label{2mom}
\eeq
leading to an exponential growth for $\bar E_c$,
with e-folding time $|\tau|$. The growth will stop at
$\bar E_c^s$ for $\Delta T<2qT$, while it will go on to
infinity for $\Delta T>2qT$. 
At the crossover $\Delta T=2qT$, $\bar E_c$ will grow linearly in
time,
$\bar E_c\simeq\Gamma(\Delta T)^2t/E_r$.
In all cases 
$\langle\tilde E_c^2\rangle\sim\bar E_c^2$.

\subsection{Similarity with the behavior of a thermal machine}
The process whereby part of the energy flux from thermostat $b$ to thermostat $a$ is highjacked
for $\Delta T>0$ to atom $c$, resembles that of work production in a thermal machine. 
By identifying
$\dot W=\dot E_c$ with the exerted power, it is possible to define an efficiency
\beq
\eta=-\dot{\bar W}/\dot{\bar Q}_b\simeq 2q,
\label{eta}
\eeq
where $\dot Q_b$ is the rate of energy loss by thermostat $b$, and use has been made of 
Eq. (\ref{three-atom rule}). Except for the case $f_X$ and $q$ tend to a finite limit for 
large $E_c$
(see point i at the end of Sec. \ref{Vanishing}), $\eta$ is a finite-time efficiency, with
a time horizon fixed by either $\tau$ or 
the time required by $f_X$ to drop to zero or to become $\sim T^2\Gamma_{bath}/\Gamma$.

Indeed, for $\Delta T<2qT$---in which case $\bar E_c$ reaches a finite limit---$\eta$ 
does exceed the Carnot value $\eta_C=\Delta T/T$, corresponding to increasingly
smaller values of $\tau$; in the case of the linear dynamics of Eq. (\ref{f_X}),
substituting Eqs. (\ref{E^s}) and (\ref{tau}) into Eq. (\ref{eta}) yields
\beq
\eta\simeq\frac{\Delta T}{T}\Big(1-\frac{E_rT}{(\Delta T)^2\Gamma\tau}\Big)^{-1}.
\eeq

Carnot efficiency is associated with $\tau\to\infty$ and a finite mean power
$\dot{\bar W}_C>0$ [in the case of the dynamics in Eq. (\ref{f_X}),
$\dot{\bar W}_C=\Gamma\Delta T^2/E_r$].  One could bypass the constraint
$\Gamma_{bath}\gg\Gamma f_X/T^2$ by sending artificially $\Gamma_{bath}\to\infty$ for finite
$\Gamma$, thus realizing a system with an infinite time horizon. In this case, the system would
exert work with a finite mean power at Carnot efficiency
at the price of diverging work fluctuations
(recall that $\langle \tilde E_c^2\rangle\sim\bar E_c^2$). Thus, there is no violation of 
the thermodynamic uncertainty relations \cite{barato15,pietzonka18}, although
applying such a concept to the present microscopic context is perhaps unphysical. It is
also to be noted that the dynamics of the system cannot be mapped onto
that of a cyclic machine, as the latter would be characterized by the linear scaling
for $\langle\tilde W^2\rangle$ that would be produced by constant 
$\langle\delta W\rangle$ and $\langle(\delta W)^2\rangle$, while in the present case
$\langle\tilde E_c^2\rangle$ scales quadratically in $t$ [see Eq. (\ref{2mom})].

\section{Conclusion}
A study of stochastic transport models with microscopic reversible dynamics has
been carried out. 
The study has focused on systems without linear momentum, whose
dynamics consists of the random exchange of energy between sites in a lattice.
Particular attention has been paid to the effect of conserved quantities,
and a general formalism has been devised to analyze non-ergodic effects and
the conditions for such effects to be observable at a macroscopic scale.

As a first application, the case of a lattice governed by binary nearest-neighbour interactions
has been considered. The general
expectation is that low-dimensional systems devoid of linear momentum should not manifest
anomalous behaviors. The analysis that has been carried out shows that the statement is 
valid only in part and that a low-dimensional lattice system in contact
with thermal baths at different temperatures will develop fluctuations with a long-range component.
Contrary to the case of systems endowed with a conserved linear momentum, however,
the amplitude of the 
long-range correlation component vanishes in the continuum limit and is not associated with
increase of the conductivity with the system size.  

The general analysis shows that
unless the system is formed by isolated components, at least
three-atom interactions are required for ergodicity breakup. 
As a second application the study focused then on the dynamics of a
quasi-one-dimensional chain with three- and four-atom interactions.
The interaction parameters have been tuned in such a way that
an additional conserved 
quantity, baptized twist, is present besides energy. The new quantity describes 
the degree of microscopic asymmetry or torsion
of the spatial energy distribution in the lattice. 
Depending on the choice of parameters, it is possible to
form out of twist an infinite hierarchy of conservation laws analogous
to those for vorticity in 2D hydrodynamics. The new conservation laws do not appear robust, in
the sense that the interaction with a thermal bath is 
sufficient---except for specific couplings---to 
restore ergodicity. When this happens, the equilibrium statistics of the system is
described by the standard Gibbs distribution for the energy.

The twist returns to play a role away from equilibrium when the system is in 
contact with
thermostats at different temperatures. Analysis of a version of the system with just
one three-atom component shows that,
depending on the sign of the temperature difference, the system may either behave as
an insulator or a sort of a heat battery which can intercept and store into its body
part of the heat flux.
Depending on the choice of parameters, the system may reach a "maximum charge'', at which 
point, depending again on the parameters and the sign of the temperature difference, 
it may either turn into
an insulator or continue to conduct heat, with a behavior that could be
likened to that of a thermal diode \cite{li04}.

Energy storage in the system has strong analogies with work
production in the Brownian gyrator and similar mesoscopic systems \cite{filliger07,sinitsyn07}.
It is thus possible to introduce such concepts as 
the efficiency of the storage process, and determine the relations with the time
horizon of the process, the mean power, and the amplitude of the work fluctuations.
It is to be stressed that the operation is carried out with microscopic 
agents which are not governed by molecular noise and friction, as it is the case, instead,
with mesoscopic systems.
The result
is of some interest because it shows that it is possible to extend some of the concepts of
standard thermodynamics below the scale typically studied by stochastic thermodynamics
\cite{sekimoto,seifert12}.

\acknowledgements I wish to thank Claudio Melis for interesting and helpful discussion.

\bibliography{sample}

\appendix
\section{Treatment of $\mathbf{\emph{E}}$-dependent interactions}
\label{Treatment}
The Fokker-Planck associated with Eq. (\ref{SDE}) can be obtained in the general case in
which ${J_l}^k={J_l}^k(\E)$ from the expression of the generator of the stochastic dynamics
(see also \cite{zinn-justin}).
The action of the latter on the generic function $g=g(\E)$ can be written in the form
\beq
\dot g=\Lcal g&=&\Gamma\{[\partial_{E^k}({J_l}^kf^{lm})]
\partial_{E^j}({J_m}^jg)
\nonumber
\\ 
&+&f^{lm}\partial_{E^k}[{J_l}^k\partial_{E^j}({J_m}^j
\rho)]\}
\eeq
(limited to this section, the summation convention
over repeated covariant-contravariant indices is reinstated).
The Fokker-Planck operator $\Lcal^\dag$ is defined as the operator conjugate to $\Lcal$
from the relation
\beq
\int[\d E]\ \rho\Lcal g=\int[\d E]\ g\Lcal^\dag\rho,
\eeq
where $[\d E]=\prod_l\d E^l$.
One carries out the necessary integrations by parts,
exploiting Eq. (\ref{boundary}) to kill boundary terms,
\beq
\frac{\rho\Lcal g}{g}&\to&\Gamma{J_m}^j\partial_{E^j}\{
{J_l}^k\partial_{E^k}(\rho f^{lm})
-\rho\partial_{E^k}({J_l}^kf^{lm})
\}
\nonumber
\\
&=&\Gamma\hat\partial_{X^m}\{f^{lm}[\hat\partial_{X^l}\rho-(\partial_{E^k}{J_l}^k)\rho]\}
=\Lcal^\dag\rho,
\eeq
and obtains the Fokker-Planck equation
\beq
\dot\rho=\hat\partial_{X^m}\{f^{lm}[\hat\partial_{X^l}\rho-
(\partial_{E^k}{J_l}^k)\rho]\}.
\label{FP general}
\eeq
As an application, one can check that the second law of thermodynamics holds
for the isolated system.  The Shannon entropy of the system is
\beq
S(t)=-\int[\d E]\ \rho\ln\rho.
\label{Shannon}
\eeq
From Eq. (\ref{FP general}) one obtains
\beq
\dot S&=&-\int[\d E]\ (\Lcal^\dag \rho)\ln\rho
\nonumber
\\
&=&\Gamma\int[\d E]\ \frac{f^{lm}
[\hat\partial_{X^l}\rho-(\partial_{E^k}{J_l}^k)\rho]
\partial_{E^j}({J_m}^j\rho)
}{\rho}
\nonumber
\\
&=&\Gamma\int[\d E]\ f^{lk}\frac{\hat\partial_{X^l}\rho\,
\hat\partial_{X^m}\rho}{\rho}
\ge 0,
\label{second law microcanonical}
\eeq
where again use has been made of Eq. (\ref{boundary}).
At equilibrium, one recovers
$S_{eq}=\ln\Omega(E,\ldots)$,
where $\Omega(E,\ldots)$ is the area of $\bfOmega$, and it is assumed that
the system was initially in a state with well-defined values of the conserved quantities.
The standard expression for the microcanonical entropy is recovered when $E$ is
the only conserved quantity.

\section{Fluctuation spectrum for the simple chain}
\label{Fluctuation spectrum}
The equation for the fluctuation amplitude is obtained from either Eq. (\ref{Fokker-Planck chain})
or Eqs. (\ref{SDE chain}) and (\ref{noise chain}), and reads
\beq
(1/2)\nabla_{kl}^2\langle E_kE_l\rangle+\delta_{kl}\langle E_k[E_{k+1}+E_{k-1}]\rangle
\nonumber
\\
=
\delta_{l,k+1}\langle E_kE_{k+1}\rangle+\delta_{l,k-1}\langle E_kE_{k-1}\rangle,
\label{moment2}
\eeq
Substituting
Eq. (\ref{decomp}) into Eq. (\ref{moment2}) and exploiting the condition
$\langle\tilde E^{le}_k\tilde E^r_l\rangle=0$
yields 
\beq
C_{k,k+1}+C_{k,k-1}-C_{kk}&=0,\quad &l=k,
\label{flsp1}
\\
{E'}^2+C_{k+1,k+1}+C_{k+1,k-1}& &\ 
\nonumber
\\
+C_{k,k+2}+C_{kk}-6C_{k,k+1}&=0,
\quad &l=k+1,
\label{flsp2}
\\
\nabla_{kl}^2C_{lk}&=0,
\quad &l>k+1.
\label{flsp3}
\eeq
By exploiting the symmetry $C_{kl}=C_{lk}$, Eqs. (\ref{flsp1}) and (\ref{flsp2}) can 
be expressed in terms of components $C_{k,l>k}$. Taylor expanding around the diagonal
and exploiting again $C_{kl}=C_{lk}$, one is left with the two equations
\beq
C_{kk}-(\partial_l-\partial_k)C_{lk}|_{l=k}=0,
\\
2{E'}^2-4C_{kk}+(\partial_l-\partial_k)C_{lk}|_{l=k}=0,
\eeq
from which one obtains the boundary condition on the diagonal
\beq
C_{kk}=\frac{2}{3}{E'}^2,
\label{BCdiag1}
\eeq
which is Eq. (\ref{BCdiag}).
Identify atoms at sites $k=0$ and $k=N$ with atoms in the respective thermal baths,
in such a way that $\bar E_0\equiv T_0$, $\bar E_N\equiv T_N$ and 
$\bar E'=(T_N-T_0)/N$.
Consistent with a hypothesis of zero correlation between atoms in the chain and in 
the bath, it is then natural to impose
\beq
C_{0k}=C_{lN}=0,
\label{BCendchain}
\eeq
Equation (\ref{flsp3}) can then be numerically solved with the Dirichlet boundary 
conditions Eqs. (\ref{BCdiag}) and (\ref{BCendchain}),
and the result is shown in Fig. \ref{microfig3}.

\end{document}